\documentclass[conference]{IEEEtran}
\IEEEoverridecommandlockouts
\usepackage{cite}
\usepackage{amsmath,amssymb,amsfonts}
\usepackage{algorithmic}
\usepackage{graphicx}
\usepackage{textcomp}
\usepackage{xcolor}
\usepackage{subcaption} 
\usepackage{hyperref}

\newcommand{\nix}[1]{}

\hypersetup{
    colorlinks=true,
    linkcolor=blue,
    filecolor=magenta,
    urlcolor=cyan,
    pdftitle={Overleaf Example},
    pdfpagemode=FullScreen,
    }

\def\BibTeX{{\rm B\kern-.05em{\sc i\kern-.025em b}\kern-.08em
    T\kern-.1667em\lower.7ex\hbox{E}\kern-.125emX}}
\begin{document}

\title{Diagnosis of Malignant Lymphoma Cancer Using Hybrid Optimized Techniques Based on Dense Neural Networks}
\author{ 

\author{\IEEEauthorblockN{Salah A. Aly}
\IEEEauthorblockA{\textit{Computing \& Data Science College} \\
\textit{Badya University}\\
October $6^{th}$ City, Giza, Egypt \\
}
\and
\IEEEauthorblockN{Ali Bakhiet}
\IEEEauthorblockA{\textit{Computer Science Department} \\
\textit{Culture \& Science October $6^{th}$ City}\\
October $6^{th}$ City, Giza, Egypt \\
}
\and
\IEEEauthorblockN{Mazen Balat}
\IEEEauthorblockA{\textit{Computer Science Department} \\
\textit{E-JUST}\\
Alexandria, Egypt\\
}
}
}

\maketitle

\begin{abstract}
Lymphoma diagnosis, particularly distinguishing between subtypes, is critical for effective treatment but remains challenging due to the subtle morphological differences in histopathological images. This study presents a novel hybrid deep learning framework that combines DenseNet201 for feature extraction with a Dense Neural Network (DNN) for classification, optimized using the Harris Hawks Optimization (HHO) algorithm. The model was trained on a dataset of 15,000 biopsy images, spanning three lymphoma subtypes: Chronic Lymphocytic Leukemia (CLL), Follicular Lymphoma (FL), and Mantle Cell Lymphoma (MCL). Our approach achieved a testing accuracy of 99.33\%, demonstrating significant improvements in both accuracy and model interpretability. Comprehensive evaluation using precision, recall, F1-score, and ROC-AUC underscores the model's robustness and potential for clinical adoption. This framework offers a scalable solution for improving diagnostic accuracy and efficiency in oncology.
\end{abstract}


\section{Introduction}
Accurate diagnosis of malignant lymphomas is critical for determining appropriate treatment strategies and improving patient outcomes \cite{gonccalves2023core}. Histopathological image analysis plays a pivotal role in this diagnostic process, where pathologists examine stained tissue sections to identify cancerous subtypes. However, the manual interpretation of these images is time-consuming and prone to variability due to the subtle morphological differences between subtypes. This makes automated, accurate, and reliable classification systems highly desirable in clinical practice.
Lymphoma diagnosis through histopathological images presents multiple challenges. The visual complexity of the images, with subtle differences between subtypes, makes accurate classification difficult \cite{kumar2023crccn}. Additionally, variability in staining techniques and tissue preparation across different laboratories adds further complexity to training AI models \cite{bai2023deep}. The limited availability of annotated data, particularly for rare lymphoma subtypes, restricts model training \cite{patricio2023challenges}. Furthermore, the lack of interpretability in AI-driven models raises concerns about their clinical adoption, as clinicians need transparency in how decisions are made \cite{wysocki2023assessing}.

The primary clinical objectives in lymphoma diagnosis are early detection, accurate subtype classification, and minimizing diagnostic variability \cite{zhang2024minimal}. Early detection is crucial for initiating timely treatment, which can significantly improve patient outcomes. Accurately distinguishing between subtypes such as CLL, FL, and MCL is essential for selecting the appropriate treatment protocol \cite{robak2023atypical}. Ensuring consistency and reproducibility in diagnoses across different clinical settings helps reduce errors and improve the quality of patient care \cite{levman2023error}. Ultimately, these objectives aim to enhance prognosis and treatment outcomes for patients with lymphoma.

This research addresses the clinical challenges of lymphoma diagnosis through the following key contributions:
\begin{enumerate}
    \item DenseNet201 was utilized for transfer learning to extract robust features for lymphoma subtype classification. Various freezing strategies were employed to balance pre-trained knowledge with domain-specific learning, resulting in improved accuracy and addressing the challenge of limited data.
    \item Harris Hawks Optimization (HHO) was applied to optimize the Dense Neural Network (DNN). This metaheuristic approach led to enhanced model accuracy, improved efficiency, and reduced overfitting, making it suitable for medical imaging tasks.
    \item A comprehensive performance evaluation was conducted using metrics such as precision, recall, F1-score, and ROC-AUC. This ensured the model’s robustness and its ability to generalize effectively across diverse clinical datasets with real-world variability.
    \item A scalable framework for automated lymphoma diagnosis was proposed. The integration of DenseNet201 and HHO created a flexible model, with potential for extension to other cancers. 
\end{enumerate}

The paper is structured as follows: Section~\ref{sec:relatedwork} provides an overview of related work in lymphoma classification. Section 3 details the materials and methods, including dataset description, data preprocessing, feature extraction using DenseNet201, and the proposed classification model. Section 4 explains the optimization process using Harris Hawks Optimization (HHO). Section 5 presents the performance evaluation metrics and results, with a detailed comparison of model performance before and after optimization.

\section{Related Work}\label{sec:relatedwork}

The classification of lymphoma subtypes using machine learning and deep learning techniques has been a significant area of research in recent years. Various studies have explored different methodologies, datasets, and models to enhance the accuracy and efficiency of lymphoma classification. This section provides an overview of key contributions in this field, highlighting the use of Artificial Neural Networks (ANNs), Convolutional Neural Networks (CNNs), Evolutionary Algorithms (EAs), transfer learning, and ensemble approaches.

Walsh et al. \cite{walsh2021evolution} investigated the effectiveness of Artificial Neural Networks (ANNs) and Deep Learning for Lymphoma classification, employing TensorFlow and Keras for network construction. They introduced a novel framework utilizing Evolutionary Algorithms (EAs) to optimize network weights. The study utilized a Convolutional Neural Network (CNN), achieving a tenfold cross-validation test accuracy of 95.64\% and a best single run test accuracy of 98.41\%. These results indicate that ANNs, when optimized with EAs, can surpass the diagnostic accuracy of the average human pathologist in classifying Lymphoma biopsies.

Rajadurai et al. \cite{rajadurai2024precisionlymphonet} utilized two Kaggle datasets for their research: a smaller dataset with 374 TIF-formatted samples (109 CLL, 124 FL, 109 MCL) and a larger dataset consisting of 15,000 images (5,000 for each subtype: CLL, FL, and MCL). The methodology involved transfer learning with pre-trained CNN models, including VGG16, VGG19, DenseNet201, InceptionV3, and Xception, alongside a stacked ensemble approach that combined InceptionV3 and Xception. For the smaller dataset, DenseNet201, InceptionV3, and Xception models achieved over 90\% accuracy, with the ensemble model reaching 97\%. For the larger dataset, the ensemble model achieved a remarkable 99\% accuracy.

In their study, {\"O}zg{\"u}r et al. \cite{ozgur2024new} utilized a dataset consisting of histopathological images of various lymphomas, including chronic lymphocytic leukemia (CLL), follicular lymphoma (FL), and mantle cell lymphoma (MCL). The methodology involved feature extraction using the GLCM method and transfer learning architectures, with principal component analysis applied for feature selection and dimensionality reduction. For classification, they employed machine learning algorithms like random forests, k-nearest neighbors (KNN), naive Bayes, and decision trees, along with deep learning models including VGG16, ResNet50, and DenseNet201. The results showed that the highest accuracy in binary classification was 94\% for CLL and FL using DenseNet201, while the lowest accuracy in binary classification was 49\% for MCL and CLL. In triple classification, the highest accuracy achieved was 82\%, with the KNN algorithm yielding the lowest performance at 36\%.

Habijan et al. \cite{habijan2024ensemble} explored the classification of lymphoma types using deep learning models. The study utilized medical imaging datasets for three common lymphoma types: chronic lymphocytic leukemia (CLL), follicular lymphoma (FL), and mantle cell lymphoma (MCL). The authors applied transfer learning on pre-trained CNN models, including VGG-19, DenseNet201, MobileNetV3, and ResNet50V2, adapting them for the lymphoma classification task. DenseNet201 achieved the highest accuracy of 98.04\%, while ResNet50V2, MobileNetV3, and VGG-19 followed with 90.13\%, 89.07\%, and 87.11\% respectively. An ensemble approach combining all four models further improved performance, reaching an accuracy of 98.89\%.

While these studies show progress, challenges like hyperparameter optimization, data limitations, and model generalizability persist. This work addresses these by applying Harris Hawks Optimization (HHO), utilizing DenseNet201 with freezing strategies, and ensuring robust performance evaluation across lymphoma subtypes.

\section{Materials and Methods Involved}

This section presents the materials and methods for developing the lymphoma classification model. It includes a histopathological image dataset, preprocessing steps, transfer learning with DenseNet201 for feature extraction, and classification using a Dense Neural Network (DNN). Hyperparameter tuning is optimized with Harris Hawks Optimization (HHO), and the model's performance is evaluated using various metrics to ensure accuracy and generalizability.

\subsection{Dataset Description}
The dataset used in this study comprises 15,000 biopsy images of malignant lymphoma \cite{orlov2010automatic}, a type of cancer affecting the lymph nodes. It specifically includes 5,000 samples for each of the three distinct types of malignant lymphoma:

\begin{itemize}
    \item \textbf{Chronic Lymphocytic Leukemia (CLL):} A slow-progressing cancer where abnormal white blood cells accumulate in the blood, bone marrow, and lymphatic tissues.
    \item \textbf{Follicular Lymphoma (FL):} A type of non-Hodgkin lymphoma that originates from the lymph nodes and has a characteristic growth pattern resembling follicles.
    \item \textbf{Mantle Cell Lymphoma (MCL):} A more aggressive form of non-Hodgkin lymphoma that arises from cells originating in the ``mantle zone'' of lymph nodes.
\end{itemize}

\begin{figure}[h]
    \centering
    \begin{subfigure}[t]{0.15\textwidth} 
        \includegraphics[width=\textwidth]{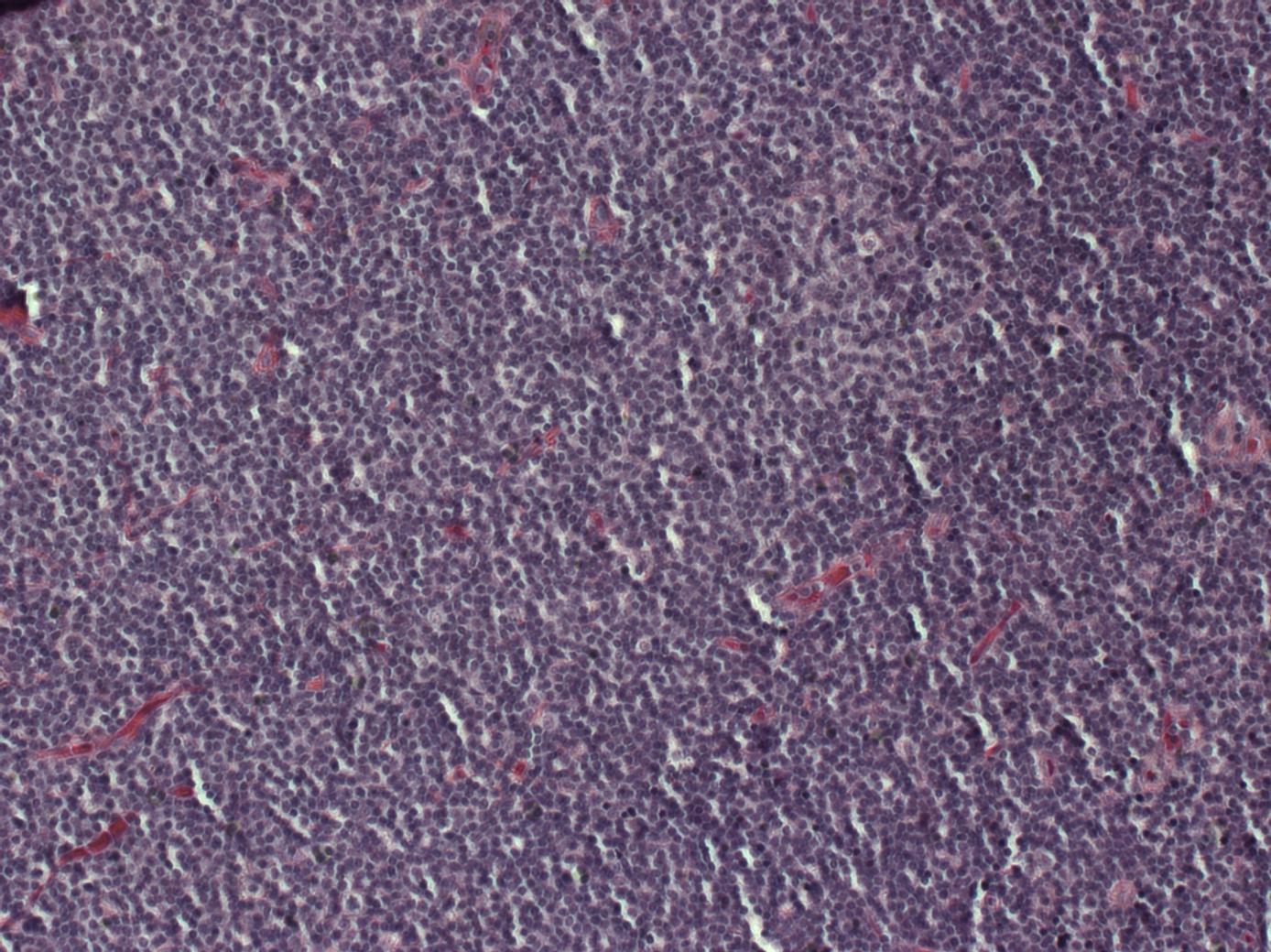}
        \caption{CLL}
        \label{fig:cll}
    \end{subfigure}
    \hfill
    \begin{subfigure}[t]{0.15\textwidth} 
        \includegraphics[width=\textwidth]{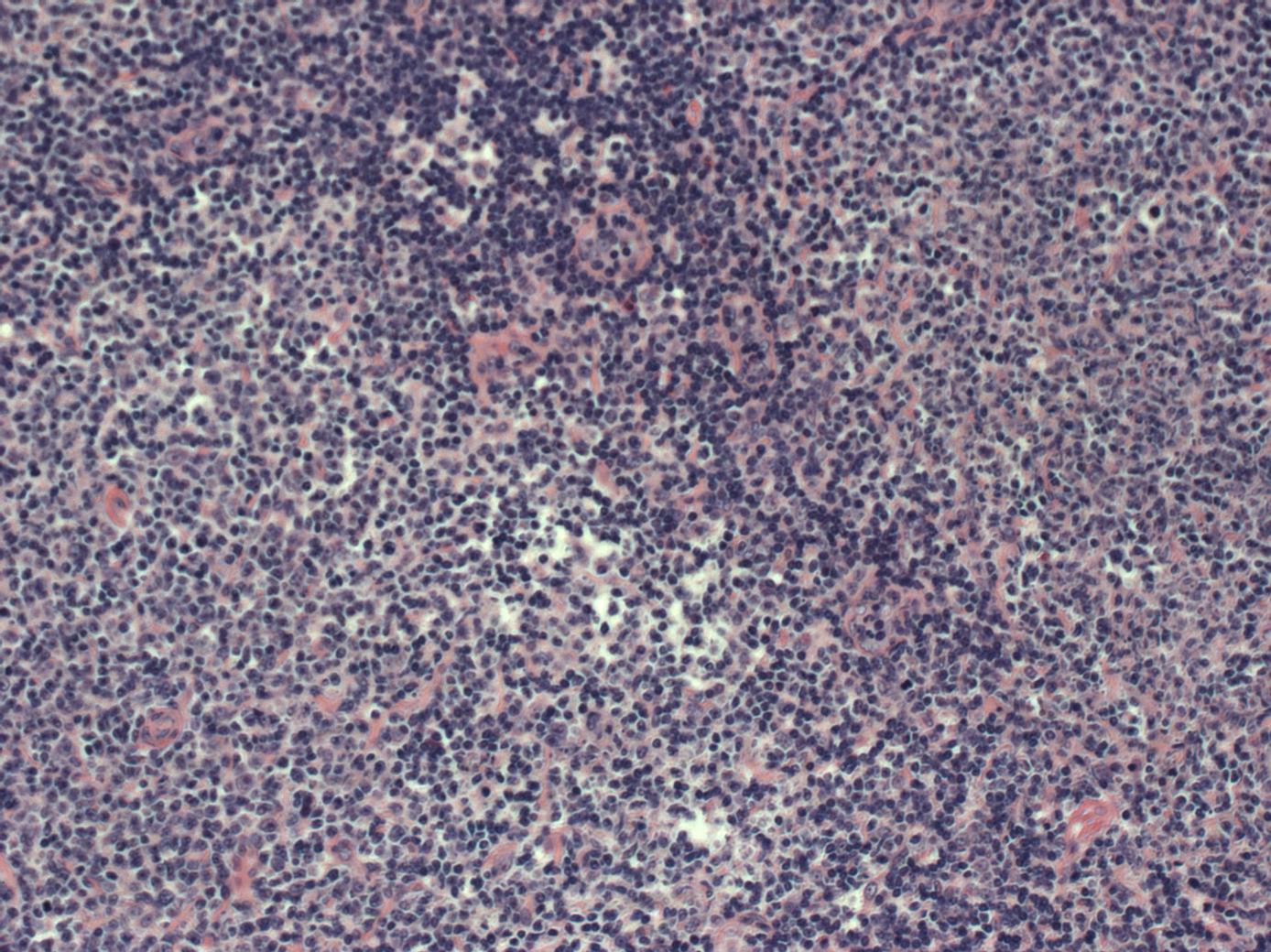}
        \caption{FL}
        \label{fig:fl}
    \end{subfigure}
    \hfill
    \begin{subfigure}[t]{0.15\textwidth} 
        \includegraphics[width=\textwidth]{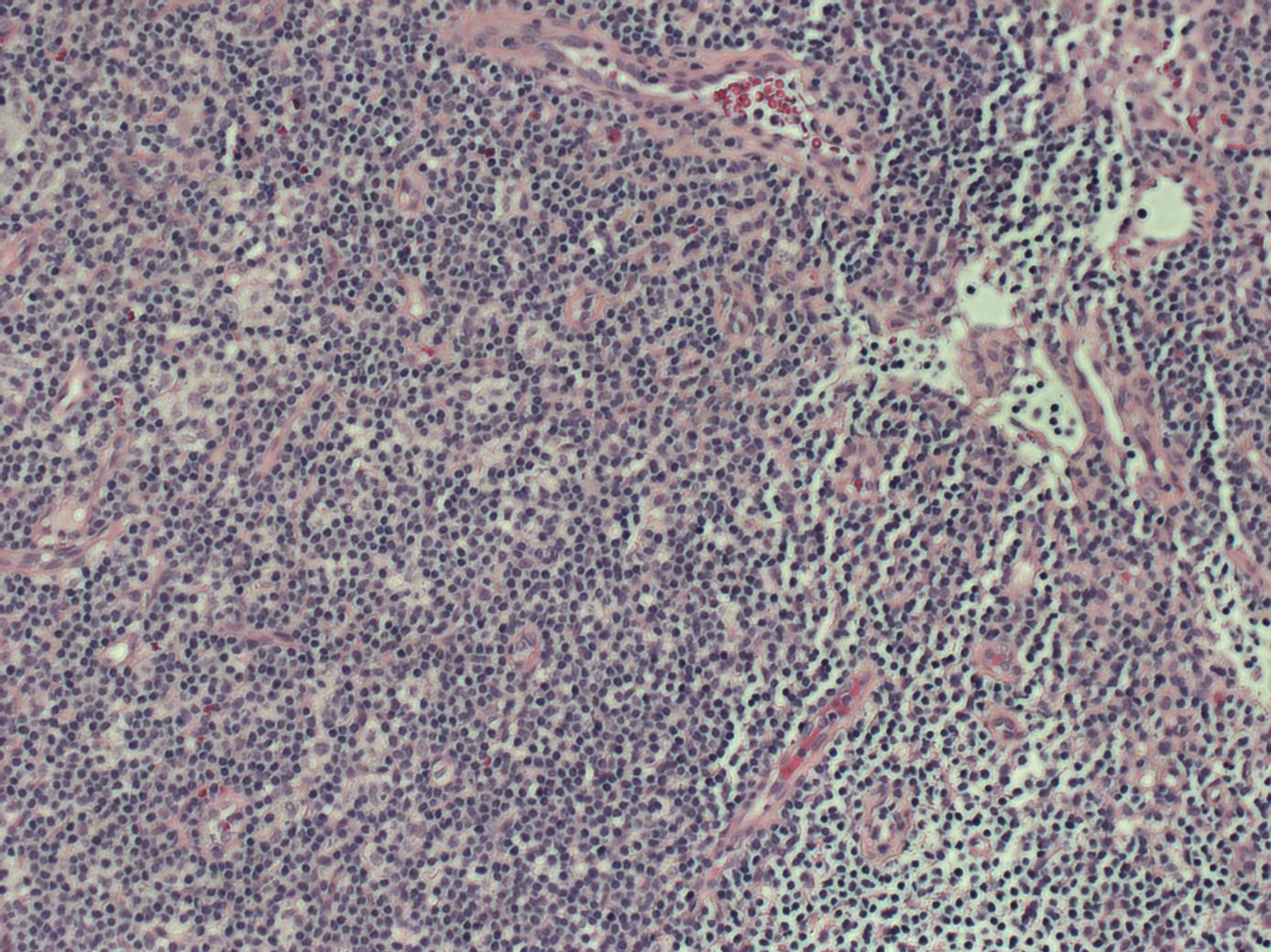}
        \caption{MCL}
        \label{fig:mcl}
    \end{subfigure}
    \caption{Representative biopsy samples for each class of malignant lymphoma: CLL, FL, and MCL.}
    \label{fig:lymphoma_samples}
\end{figure}

These samples were prepared by various pathologists across different sites, which introduces a significant amount of staining variation in the Hematoxylin and Eosin (H\&E) stained biopsies. This variation mirrors the real-world challenges of histopathological diagnosis, where differences in sample preparation can affect the consistency of visual interpretation (see Figure~\ref{fig:lymphoma_samples}).

The ability to distinguish between these types of lymphoma using H\&E-stained biopsies is critical for accurate diagnosis and treatment planning. However, this task is typically complex, requiring the expertise of highly specialized pathologists. The current standard involves the use of class-specific molecular probes, which, while reliable, are resource-intensive and time-consuming.

This dataset provides an opportunity to develop and test automated methods for lymphoma classification that could potentially reduce the workload on pathologists and improve diagnostic consistency. By including samples with inherent staining variation, the dataset allows for the evaluation of model robustness in realistic clinical settings.

The dataset is a curated collection of biopsy images that reflect the natural diversity seen in clinical practice. Samples were sectioned and stained with H\&E by different pathologists, reflecting the staining variability encountered across different laboratories and institutions. This variability is critical for developing generalizable diagnostic models.

\subsection{Data Preprocessing}

The histopathological images, originating from various medical datasets, were resized to a fixed resolution of 224 × 224 pixels. This step was necessary to standardize the image dimensions, as the DenseNet201 model requires this specific input size. By resizing the images, we ensured uniformity across the dataset, which is critical for the consistency of the feature extraction process. This step not only facilitates computational efficiency but also helps prevent distortions or scaling issues that could arise from varying image sizes.

\subsection{Feature Extraction}

\subsubsection{Transfer Learning Using DenseNet201}

For feature extraction, we leveraged the DenseNet201 architecture \cite{huang2017densely}, pre-trained on the ImageNet dataset. DenseNet201 is known for its deep and dense connections between layers, which enable it to efficiently capture intricate, hierarchical features. This ability is particularly useful for differentiating between subtle variations in lymphoma subtypes, as the model can effectively extract both low-level (e.g., textures and shapes) and high-level semantic information (e.g., tissue patterns and cellular structures) \cite{jiang2023deep}.

\subsubsection{Freezing Strategies}
To balance the transfer learning capabilities of DenseNet201 with the need for task-specific adaptation, we implemented three distinct freezing strategies. These strategies control how much of the pre-trained network is kept static (frozen) and how much is fine-tuned during training for lymphoma classification \cite{iman2023review}:

\begin{itemize}
    \item \textbf{Total Freeze}: All layers of DenseNet201 were frozen, retaining the pre-trained weights without any updates during training. This approach preserves the knowledge learned from the ImageNet dataset entirely, which can be advantageous when the target domain has similar features.

    \item \textbf{Half Freeze}: The first half of the DenseNet201 layers were frozen, while the remaining layers were left trainable. This setup allows the model to maintain robust feature extraction capabilities in its early layers while fine-tuning deeper layers for the nuances specific to lymphoma classification.

    \item \textbf{Last Block Freeze}: All layers except the final block were frozen. The last block, consisting of the most abstract and high-level layers, was trainable, allowing the model to adjust to the particular characteristics of the target dataset while still benefiting from the lower-level features extracted by the pre-trained layers.
\end{itemize}

\subsubsection{Feature Vector Concatenation}
After configuring the DenseNet201 models with the aforementioned freezing strategies, we extracted feature vectors from each configuration. These vectors were then concatenated to form a comprehensive, high-dimensional feature vector. By merging the outputs from models with different freezing strategies, we encapsulated a wide variety of representations—ranging from low-level structural features to high-level semantic insights—ensuring that the classification model receives a rich and diverse set of input features \cite{khanday2021taxonomy}.

\subsection{Classification Models}

The classification of lymphoma subtypes was handled by a Dense Neural Network (DNN) \cite{mei2020densely}. The DNN was specifically designed to process the concatenated feature vectors and classify them into one of the three lymphoma subtypes: chronic lymphocytic leukemia (CLL), follicular lymphoma (FL), and mantle cell lymphoma (MCL).

\paragraph{Model Architecture}
The deep neural network (DNN) model used in this study is composed of several key components, as summarized in Table \ref{tab:model_architecture}.

\begin{table}[h!]
\centering
\caption{Summary of the DNN Architecture Components}
\label{tab:model_architecture}
\begin{tabular}{|c|c|}
\hline
\textbf{Component} & \textbf{Description} \\
\hline
Input Layer & 512 Neurons (DenseNet201 features) \\
\hline
Dropout Layer & 50\% Dropout rate to prevent overfitting \\
\hline
Hidden Layers & 3 Layers: 256, 128, 64 Neurons (ReLU activation) \\
\hline
Batch Normalization & Applied after each hidden layer \\
\hline
Output Layer & 3 Neurons (Softmax for class probabilities) \\
\hline
\end{tabular}
\end{table}

The input layer contains 512 neurons, matching the size of the concatenated feature vector from DenseNet201 models. A 50\% dropout layer follows to reduce overfitting by randomly deactivating neurons during training, which encourages learning generalized features and improves performance on unseen data.

Next, the network includes three fully connected hidden layers with 256, 128, and 64 neurons, respectively. Each layer uses the ReLU activation function to introduce non-linearity, helping the model capture complex data patterns. Batch normalization is applied after each hidden layer to stabilize training and reduce internal covariate shift, enhancing robustness.

Finally, the output layer has three neurons corresponding to the three lymphoma subtypes, with a softmax activation function providing class probabilities \cite{pearce2021understanding}.

\paragraph{Model Compilation}
The DNN was compiled using the settings shown in Table \ref{tab:model_compilation}.

\begin{table}[h!]
\centering
\caption{DNN Compilation Configurations}
\label{tab:model_compilation}
\begin{tabular}{|c|c|}
\hline
\textbf{Component} & \textbf{Description} \\
\hline
Optimizer & Adam \cite{liu2021adam} \\
\hline
Loss Function & Categorical cross-entropy \cite{li2022improved} \\
\hline
Learning Rate Adjustment & ReduceLROnPlateau \cite{thakur2024transformative} \\
\hline
\end{tabular}
\end{table}

The Adam optimizer was selected for its adaptability in adjusting learning rates during training \cite{liu2021adam}. Categorical cross-entropy, a standard for multi-class classification, was chosen as the loss function \cite{li2022improved}. Additionally, the ReduceLROnPlateau callback was used to adjust the learning rate when validation accuracy plateaued, promoting efficient convergence \cite{thakur2024transformative}.

\subsection{Optimized Dense Neural Network (ODNN) with Harris Hawks Optimization (HHO)}

To enhance the classification performance further, we applied the Harris Hawks Optimization (HHO) algorithm to optimize the hyperparameters of the DNN \cite{shehab2022harris}. HHO is a metaheuristic optimization algorithm inspired by the cooperative hunting strategies of Harris hawks, which balance exploration and exploitation during the search process \cite{hussien2022self}.

The optimization process involved the following steps:

\begin{itemize}
    \item \textbf{Fitness Function}: A fitness function was designed to evaluate different configurations of the DNN based on classification accuracy and loss. The fitness function provided a quantitative measure of each configuration's performance, guiding the HHO algorithm towards the best model setup.

    \item \textbf{Hyperparameter Search}: The HHO algorithm explored the hyperparameter space, searching for optimal values such as the number of neurons in each layer, learning rate, dropout rate, and batch size. By iterating over different configurations, HHO fine-tuned the DNN to maximize classification accuracy.

    \item \textbf{Convergence Monitoring}: Throughout the optimization process, we monitored accuracy and loss plots to ensure the model was converging as expected. These plots provided insights into the performance of the ODNN as the HHO algorithm refined the hyperparameters.
\end{itemize}

\subsection{Performance Evaluation Metrics}

To assess the classification model for malignant lymphoma subtypes, we employed a set of metrics that evaluate both overall accuracy and detailed classification performance across the training and testing phases.

\subsubsection{Accuracy}
Accuracy represents the proportion of correctly classified instances in both training and testing datasets.

\begin{equation}
\text{Accuracy} = \frac{TP + TN}{TP + TN + FP + FN}
\end{equation}

Where:
\begin{itemize}
    \item[ ]  $TP$ = True Positives, $TN$ = True Negatives
    \item[ ]  $FP$ = False Positives, $FN$ = False Negatives
\end{itemize}

\subsubsection{Precision}
Precision calculates the accuracy of positive predictions.

\begin{equation}
\text{Precision} = \frac{TP}{TP + FP}
\end{equation}

\subsubsection{Recall}
Recall, or sensitivity, measures the proportion of actual positives correctly identified.

\begin{equation}
\text{Recall} = \frac{TP}{TP + FN}
\end{equation}

\subsubsection{F1-Score}
The F1-Score balances precision and recall.

\begin{equation}
\text{F1-Score} = 2 \times \frac{\text{Precision} \times \text{Recall}}{\text{Precision} + \text{Recall}}
\end{equation}

\subsubsection{Cohen's Kappa Score}
Cohen's Kappa evaluates the agreement between predicted and actual classifications, adjusting for chance.

\begin{equation}
\kappa = \frac{p_o - p_e}{1 - p_e}
\end{equation}

Where $p_o$ is the observed agreement, and $p_e$ is the expected agreement by chance.

\subsubsection{Confusion Matrix}
The confusion matrix summarizes the classification results as:

\begin{equation}
\begin{bmatrix}
TP & FP \\
FN & TN
\end{bmatrix}
\end{equation}

\medskip

\subsubsection{ROC-AUC}
The ROC-AUC quantifies the model's ability to distinguish between classes by plotting the true positive rate (Recall) against the false positive rate (FPR), with the area under the curve (AUC) defined as:

\begin{equation}
\text{AUC} = \int_{0}^{1} \text{TPR}(t) \, d\text{FPR}(t)
\end{equation}

Where:
\begin{itemize}
    \item[ ]  TPR = True Positive Rate, FPR = False Positive Rate
\end{itemize}

These metrics offer a comprehensive evaluation of the model's performance, from overall accuracy to detailed classification behavior. \\

\begin{figure}[t]
    \centering
    \includegraphics[width=0.5\textwidth]{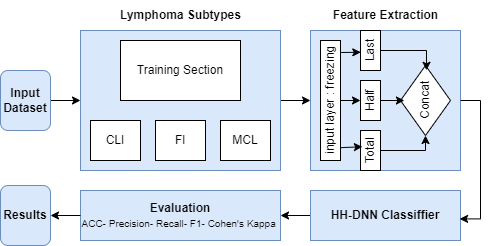}
    \caption{An overview of the methodology for lymphoma classification}
    \label{fig:methodology_flow}
\end{figure}

This methodology offers a detailed yet streamlined approach to tackling the complex task of lymphoma subtype classification, utilizing state-of-the-art techniques in transfer learning, neural network design, and hyperparameter optimization. By combining multiple freezing strategies with metaheuristic optimization, as summarized in Figure~\ref{fig:methodology_flow}, we ensure that the model is both adaptable and highly accurate.

\medskip

\section{Performance Results}

This section presents the performance comparison of the Dense Neural Network (DNN) before and after optimization using the Harris Hawks Optimization (HHO) algorithm. The results demonstrate improvements in various performance metrics post-optimization, and we also compare our results with key studies in the field.

Table \ref{tab:results_comparison} summarizes the key performance metrics for the DNN and the optimized DNN (ODNN) after applying HHO. The optimization led to notable improvements in accuracy, precision, recall, and other classification metrics.

\begin{table}[h]
    \centering
    \caption{Performance Metrics Comparison: DNN vs ODNN (with HHO)}
    \label{tab:results_comparison}
    \begin{tabular}{|l|c|c|}
        \hline
       ~~ \textbf{Metric} & \textbf{DNN} & \textbf{ODNN with HHO} \\
        \hline
        Training Accuracy & 97.38\% & 99.95\% \\
        Testing Accuracy & 97.56\% & 99.33\% \\
        Precision (Class 0: FL) & 97.60\% & 99.45\% \\
        Precision (Class 1: MCL) & 96.04\% & 99.00\% \\
        Precision (Class 2: CLL) & 100.00\% & 100.00\% \\
        Recall (All Classes) & 97.56\% & 99.33\% \\
        F1-Score (All Classes) & 97.56\% & 99.33\% \\
        Kappa Score & 96.33\% & 99.00\% \\
        ROC-AUC & 0.98 & 0.99+ \\
        Loss & 0.023 & 0.0053 \\
        \hline
    \end{tabular}
\end{table}

\medskip

The optimization resulted in a marked increase in both training and testing accuracy. Training accuracy improved from 97.38\% to 99.95\%, while testing accuracy rose from 97.56\% to 99.33\%. This demonstrates the model’s enhanced ability to generalize to unseen data. Additionally, there were significant improvements in precision, particularly for Mantle Cell Lymphoma (Class 1), where precision increased from 96.04\% to 99.00\%. The Kappa score, which measures the agreement between predicted and actual labels, improved from 96.33\% to 99.00\%, reflecting the model's higher classification reliability. The decrease in loss from 0.023 to 0.0053 further indicates better learning and optimization.

In order to further contextualize the results, Table \ref{tab:comparison_with_other_studies} provides a comparison of the performance of the ODNN (optimized with HHO) on the larger dataset of 15,000 images against results from other prominent studies in the field of lymphoma classification.

\begin{table}[h]
    \centering
    \caption{Comparison of Lymphoma Classification Results with Other Studies}
    \label{tab:comparison_with_other_studies}
    \begin{tabular}{|l|c|c|}
        \hline
        ~~ \textbf{Study} & \textbf{Method} & \textbf{Accuracy} \\
        \hline
        Walsh et al. \cite{walsh2021evolution} & CNN with EA & 98.41\% \\
        Rajadurai et al. \cite{rajadurai2024precisionlymphonet} & Ensemble (InceptionV3 + Xception) & 99.00\% \\
        {\"O}zg{\"u}r et al. \cite{ozgur2024new} & Transfer Learning + PCA & 82\% \\
        Habijan et al. \cite{habijan2024ensemble} & DenseNet201 + Ensemble & 98.89\% \\
        \textbf{Our Study} & \textbf{ODNN (with HHO)} & \textbf{99.33\%} \\
        \hline
    \end{tabular}
\end{table}

\medskip

Our optimized DNN using the Harris Hawks Optimization algorithm achieved an accuracy of 99.33\%, outperforming several other studies, particularly in terms of accuracy on the larger dataset. The optimization process, therefore, not only enhances the model's ability to generalize and perform well on unseen data, but also places it among the top-performing approaches in the field.

\begin{figure}[h]
    \centering
    \begin{subfigure}[b]{0.24\textwidth}
        \centering
        \includegraphics[width=\textwidth,height=.9\textwidth]{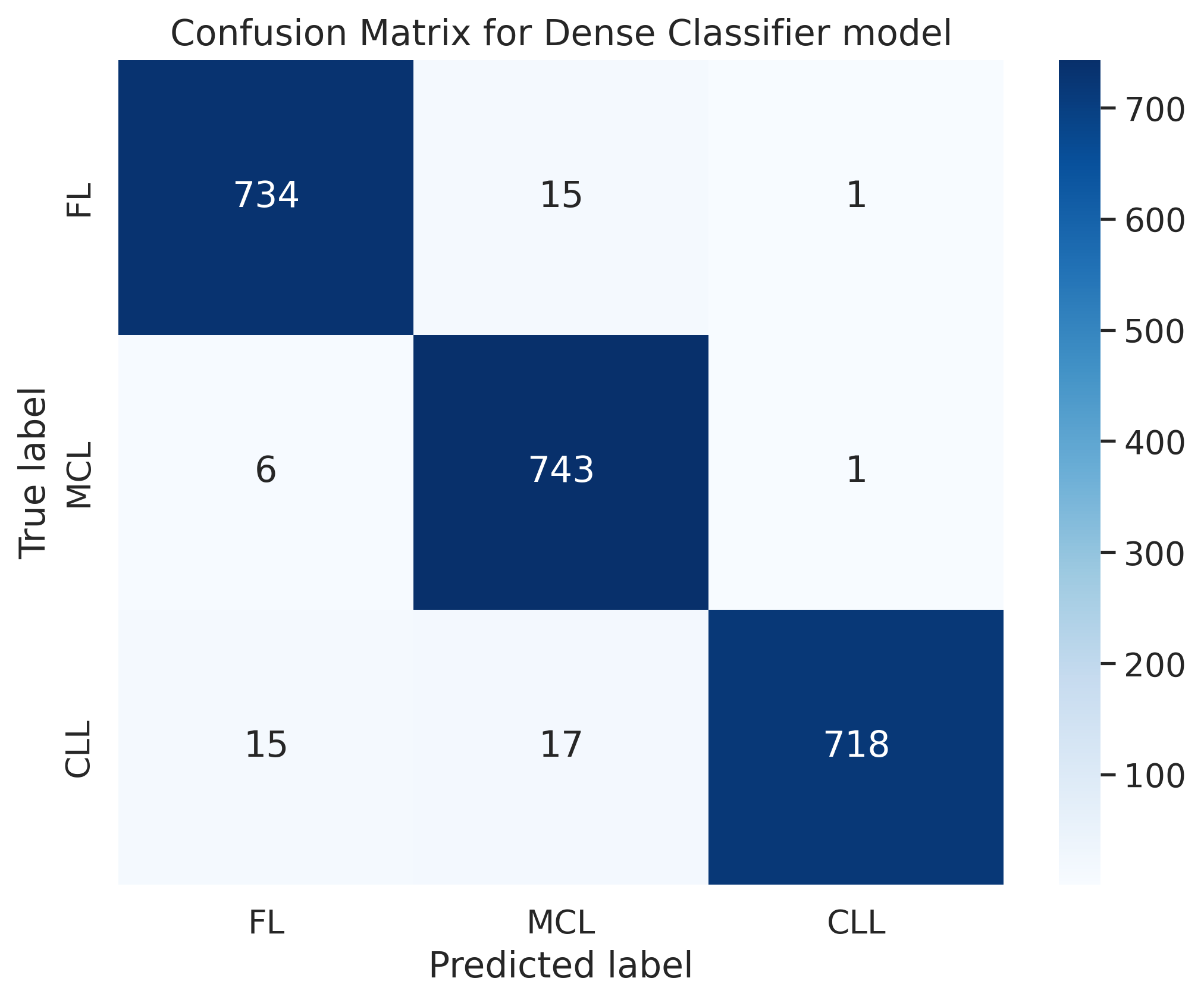}
        \caption{Confusion Matrix Before Optimization}
        \label{fig:confusion_matrix_before}
    \end{subfigure}
    \hfill
    \begin{subfigure}[b]{0.24\textwidth}
        \centering
        \includegraphics[width=\textwidth,height=.9\textwidth]{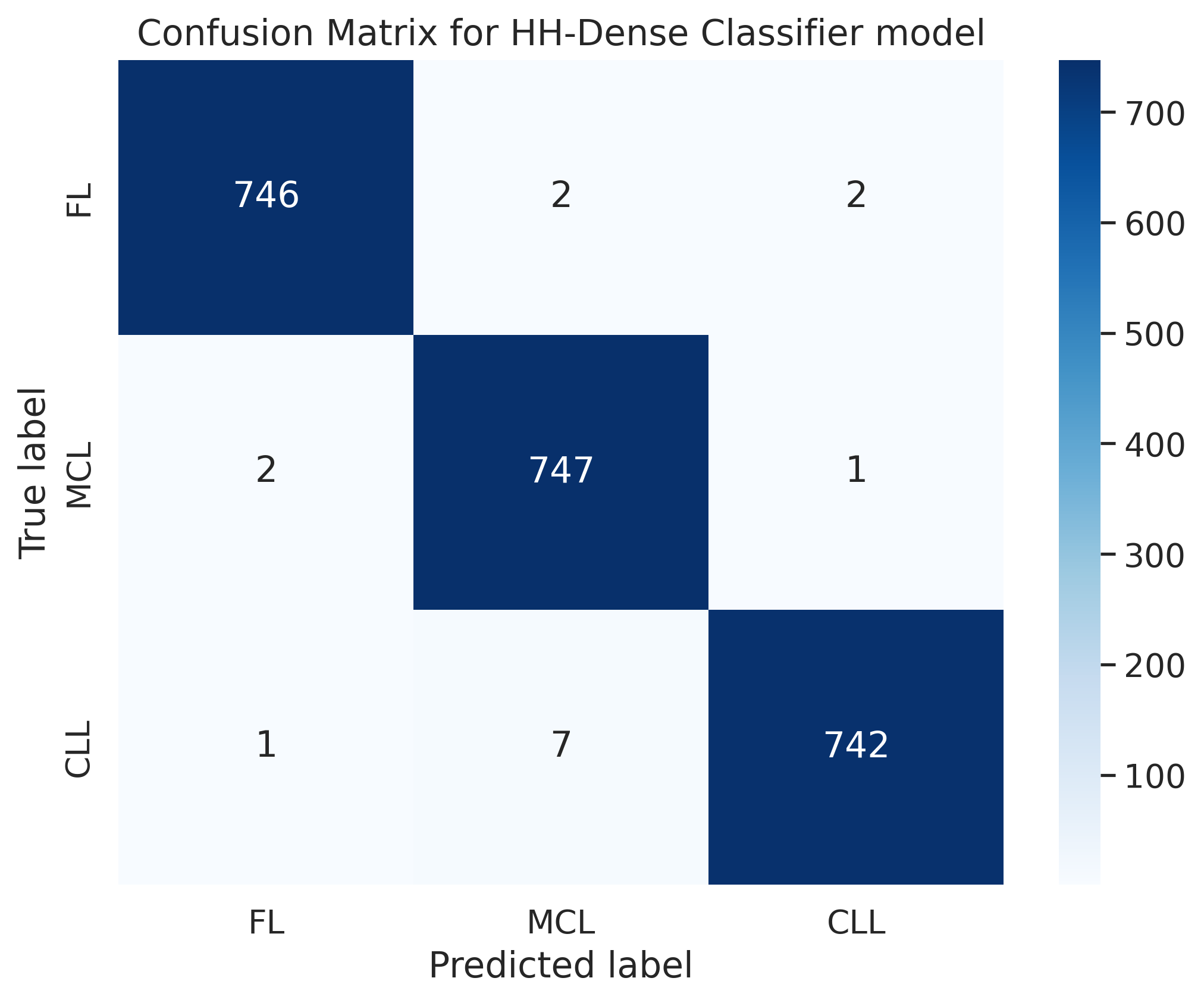}
        \caption{Confusion Matrix After Optimization}
        \label{fig:confusion_matrix_after}
    \end{subfigure}
    \caption{Confusion Matrices Before and After Optimization}
    \label{fig:confusion_matrices}
\end{figure}

The confusion matrices in Figures \ref{fig:confusion_matrix_before} and \ref{fig:confusion_matrix_after} illustrate the distribution of classification results before and after optimization. After optimization, the number of misclassifications decreased notably, particularly for follicular lymphoma and mantle cell lymphoma. Most predictions fall along the diagonal, indicating correct classifications across all classes.

\medskip

\begin{figure}[htbp]
    \centering
        \includegraphics[width=.4\textwidth]{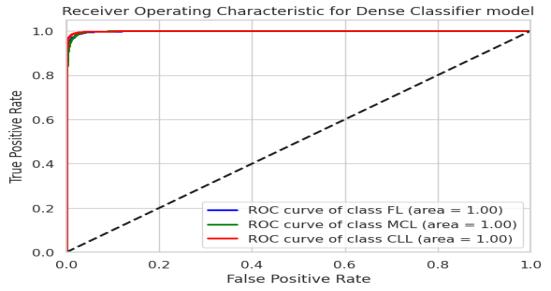}
        \caption{ROC Curve Before Optimization (DNN)}
        \label{fig:roc_before}
        \includegraphics[width=.4\textwidth]{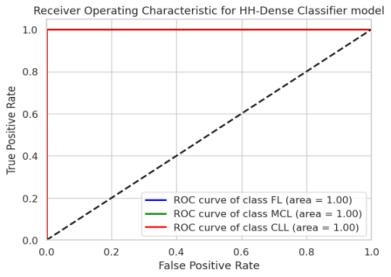}
        \caption{ROC Curve After Optimization (ODNN with HHO)}
        \label{fig:roc_after}
    \label{fig:roc_curves}
\end{figure}

The ROC curves before and after optimization (Figures \ref{fig:roc_before} and \ref{fig:roc_after}) provide additional insight into the model's performance. The ROC-AUC score, which was already high at 0.98 before optimization, improved further to over 0.99, indicating an enhanced ability to discriminate between classes. The optimized model's ROC curve approaches the ideal point, especially for Mantle Cell Lymphoma, which experienced the most significant improvement in precision and recall.

In summary, the Harris Hawks Optimization algorithm significantly improved the DNN’s overall performance. The optimized model showed higher accuracy, precision, recall, and reduced classification errors. The enhanced generalization capabilities ensure the model performs reliably on both training and testing datasets, reducing the likelihood of overfitting and increasing robustness in predicting lymphoma subtypes.

\medskip

\section{Conclusion and Future Work}

\medskip

This study presents a novel, hybrid deep learning framework for the automated diagnosis of malignant lymphoma, combining DenseNet201 for feature extraction with a Dense Neural Network (DNN) optimized using Harris Hawks Optimization (HHO). The proposed model achieved a remarkable testing accuracy of 99.33\%, along with significant improvements in precision, recall, and F1-score, particularly for challenging subtypes such as Mantle Cell Lymphoma (MCL). By addressing key challenges, such as staining variability and limited annotated data, this work showcases the potential for scalable and highly accurate diagnostic tools in clinical oncology.

Future work will focus on enhancing the interpretability of the model, ensuring that clinicians can understand and trust the automated decisions. Additionally, the framework can be expanded to other cancer types, exploring its generalizability across diverse histopathological datasets, and improving diagnostic consistency across various healthcare environments. Further integration of explainable AI techniques and real-time deployment strategies will also be pursued, promoting adoption in real-world clinical settings and improving diagnostic consistency across various healthcare environments.

\medskip
\bibliographystyle{IEEEtran}
\bibliography{IEEEabrv,cite}
\end{document}